\documentclass[a4paper]{jpconf}
\usepackage{graphicx}
\begin{document}
\title{Resolving Octant Degeneracy at LBL experiment by Combining Daya Bay Reactor Setup}

\author{Kalpana Bora$^1$, Debajyoti Dutta$^2$}

\address{Physics Department, Gauhati University, Assam, India}

\ead{$^1$kalpana@gauhati.ac.in, $^2$debajyotidutta1985@gmail.com}

\begin{abstract}
Long baseline Experiment(LBL) have promised
to be a very powerful experimental setup to study various issues
related to Neutrinos. Some ongoing and planned LBL and medium baseline experiments are-
T2K, MINOS, NOvA, LBNE, LBNO etc. But, the long baseline experiments are crippled
due to presence of some parameter degeneracies, like the Octant -
degeneracy. In this work, we first show the presence of Octant degeneracy in LBL experiments and then combine it with Daya Bay Reactor experiment at
different values of CP violation phase. We show that the Octant degeneracy
in LBNE can be resolved completely with this proposal.
\end{abstract}

\section{Introduction}
In the quest to find
the unknowns in neutrino sector, long baseline neutrino experiments (LBNEs) have an
important role to play. But there is a major drawback of the
LBNEs-the presence of parameter degeneracies \cite{1,2}.
The appearance of several disconnected region in multi-dimensional
space of oscillation parameters for a given experiment is due to the
inherent structure of three flavour neutrino oscillation probabilities.
Because of these, it is not exactly possible to pin-point, which one
is the exact (true) solution. These degeneracies can be classified
in three ways.
Appearance of two disconnected regions in the ($\delta_{cp},\theta_{13}$)
plane for $\nu_{\mu}\rightarrow\nu_{e}$ channel in neutrino and anti neutrino
mode leads to intrinsic or ($\delta_{cp},\theta_{13}$)-degeneracy \cite{3,4}. Appearance of two solutions corresponding to two sign of $\Delta{m}^{2}$
gives hierarchy or sign ($\Delta{m}^{2}$)-degeneracy \cite{5} and 
appearance of solutions corresponding to $\theta_{23}$ and $\pi/2-\theta_{23}$
at different values of $\delta_{cp}$ and $\theta_{13}$ is called octant or ($\theta_{23}$) -degeneracy \cite{6}. In literature, various methods to resolve these degeneracies have
been proposed. The use of spectral information \cite{7}, combining experiments at various baselines and/or
(L/E)-values \cite{2}, combining $\nu_{\mu}\rightarrow\nu_{e}$ and $\nu_{e}\rightarrow\nu_{\tau}$
oscillation channels \cite{8}, combining long baseline(LBL) and reactor experiments \cite{9,10} and combination of LBNE and atmospheric experiments \cite{11,12} are a few of them.\\
 Recent global fit of all neutrino oscillation parameters \cite{13}, points towards the lower octant [$\theta_{23}<\frac{\pi}{4}$], but its still not clear which octant is the true octant. Again, bounds on sterile oscillation parameters from Reactor experiments can be found on \cite{14}.\\
In this work, we have considered a LBNE like setup and combined it with Daya Bay \cite{15} reactor setup with the help of simulation using Globes \cite{16} to check what information we can get from it. We call it LBNE- like setup because LBNE collaboration is planning to reconfigure their setup with liquid argon detector. But still our work is relevant in the sense that this result can be implemented to any long baseline experiment having water cherenkov detector at a baseline $\sim 1300$ km. LBNE neutrino beamline is driven by the physics of different oscillation channels like -${\nu}_{\mu}\rightarrow{\nu}_{e}$, ${\nu}_{\mu}\rightarrow{\nu}_{\mu}$, ${\nu}_{\mu}\rightarrow{\nu}_{\tau}$ etc. We have considered the ${\nu}_{\mu}\rightarrow{\nu}_{e}$ appearance channel in this study.

 Presence of octant degeneracy, in LBNE like 
setup is confirmed by
exclusion curves in Fig.1(1st column) in the $\theta_{13}-\delta_{CP}$ plane.  Next, we combine LBNE like setup with
Daya Bay reactor experiment, to resolve the above mentioned octant
degeneracy as shown in Fig.1 (2nd column). We find that after combining,
the degenerate exclusion curves disappear. With this proposed combination, it is possible to resolve octant degeneracy. In fig.2, we have pinpointed the octant after combining LBNE like setup with Reactor setup at $\delta_{CP}$ =$0^{\circ}$. We have
done this analysis for a running time of 5 years in neutrino and 5 years of anti- neutrino. The results
have been presented for different values of $\delta_{CP}$ phase =
$0^{\circ}, 90^{\circ}, 150^{\circ}$. We have tested
these results for a wide range of $\delta_{CP}$,
as we do not know the exact value of this parameter
in leptonic sector, from experiments.
\section{ Experiment Details}
In our analysis, we have considered the LBNE to DUSEL beamline \cite{17} of 
1300 km baseline that is pointing from NuMI (Neutrino Main Injector). 
The Detector is a 300 kiloton water Cherenkov detector. Running time is
5 years for neutrinos and 5 years for anti-neutrinos. Power considered is 700 KW. Daya Bay \cite{15} has three pairs of reactors at Daya Bay and Ling Ao I, which
generate 11.6 GW of power. Daya Bay consists of three underground
experimental halls, one far and two near, linked by horizontal tunnels.
Eight identical cylindrical detectors are employed to measure the
neutrino flux. The mass of each detector is about 20 tons. Equal mass
of near and far detector will reduce the systematic errors. Four of
these eight detectors are at the far zone while two detectors are
kept in each near zone. The distance of the detectors from the reactor
cores at the Daya Bay site is 363 m while this distance at the Ling
Ao site is 481 m. The far detectors are at 1985 m and 1615 m respectively
from the Daya Bay and Ling Ao reactor sites. Energy resolution used
is 5$\%$. In this work, we have used only two pairs of reactors.
\section{Results and Analysis}

\begin{figure}[!t]
\begin{minipage}{14pc}
\includegraphics[width=18pc]{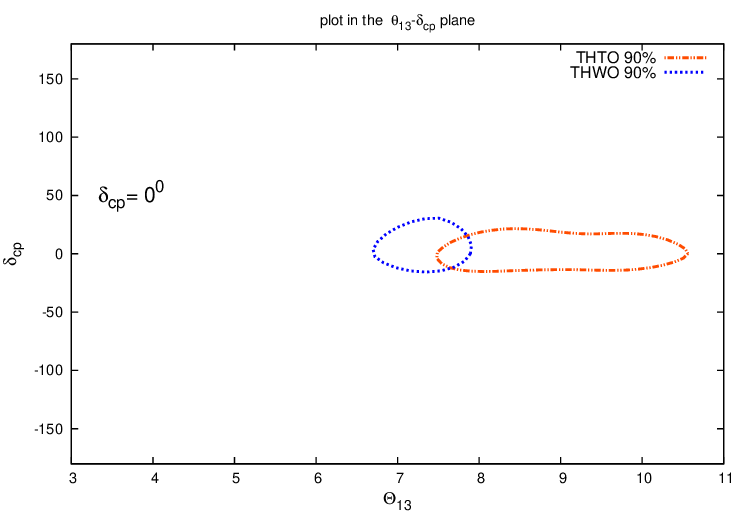}
\end{minipage}\hspace{2.5pc}%
\begin{minipage}{14pc}
\includegraphics[width=18pc]{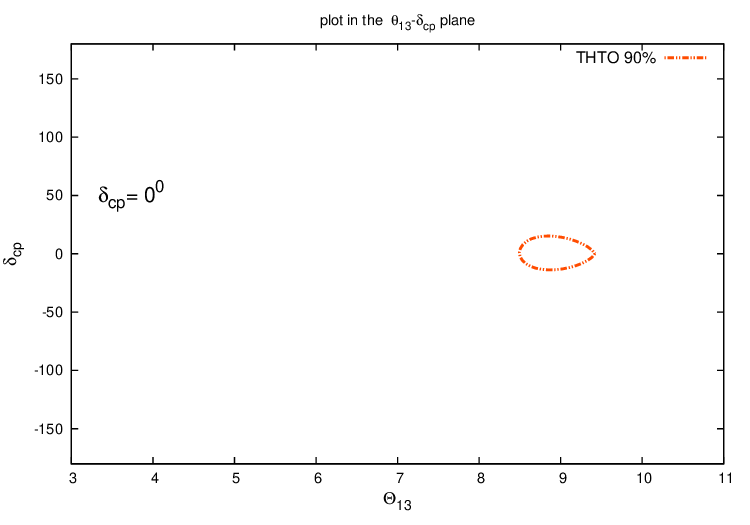}
\end{minipage} 
\begin{minipage}{14pc}
\includegraphics[width=18pc]{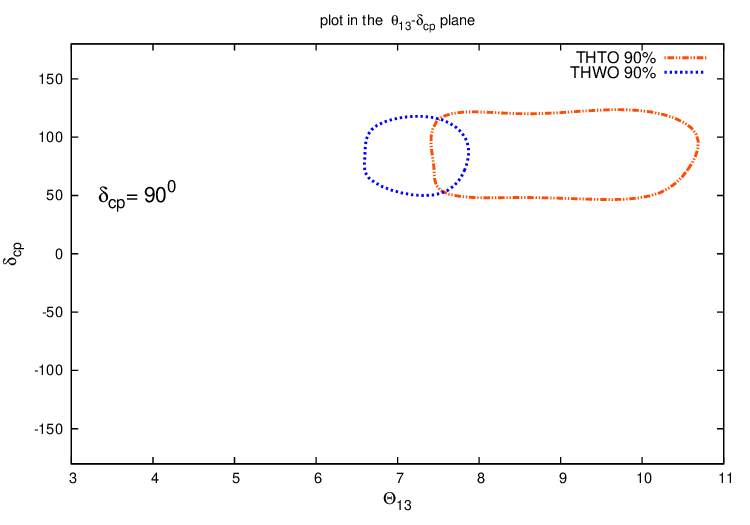}
\end{minipage}\hspace{2.5pc}%
\begin{minipage}{14pc}
\includegraphics[width=18pc]{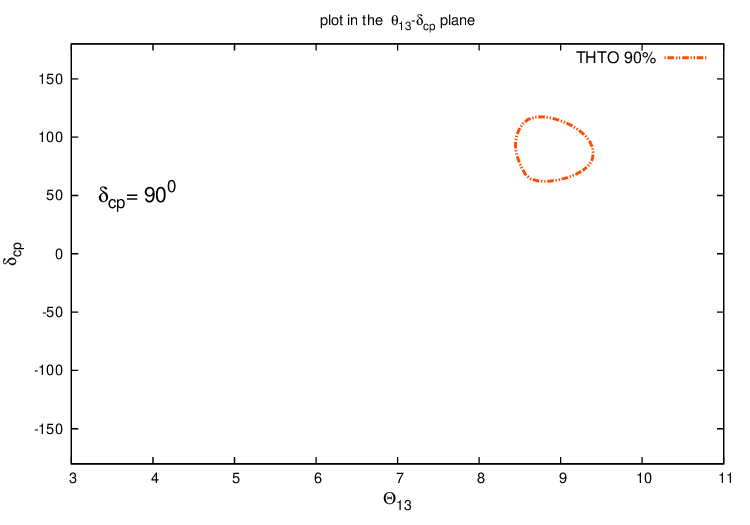}
\end{minipage} 
\begin{minipage}{14pc}
\includegraphics[width=18pc]{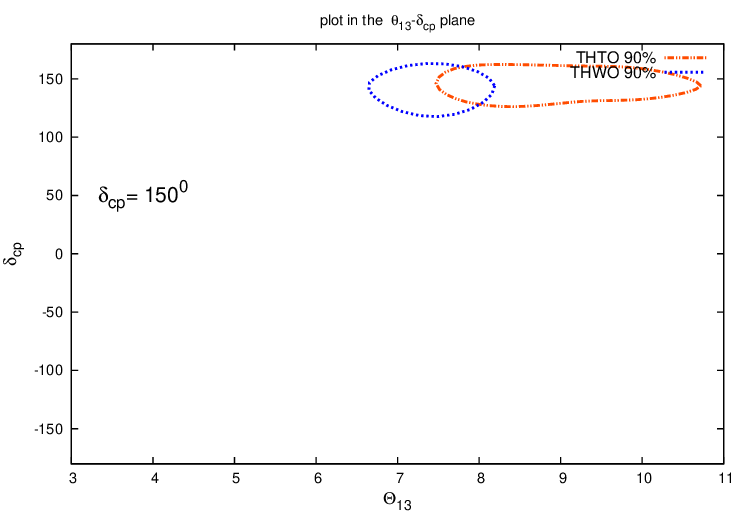}
\end{minipage}\hspace{2.5pc}%
\begin{minipage}{14pc}
\includegraphics[width=18pc]{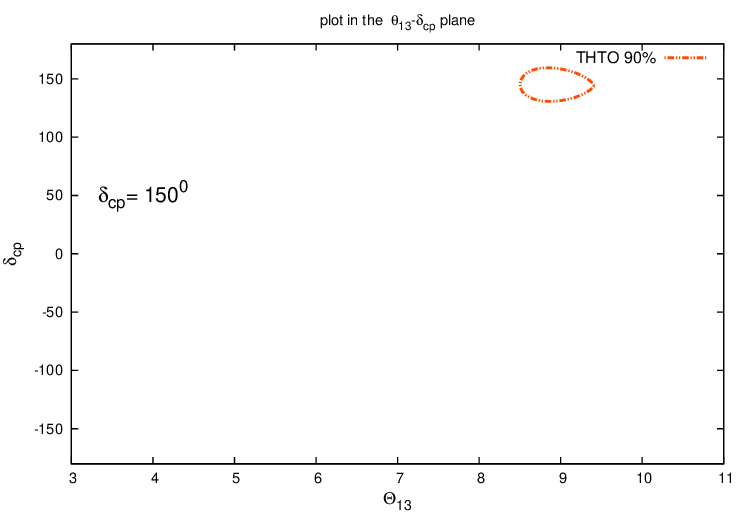}
\end{minipage} 
\caption{Contours in $\delta_{CP}-\theta_{13}$ plane for LBNE-like setup(1st column)
and LBNE-like+Daya Bay setup(2nd column) at different $\delta_{cp}$ values and at 90$\%$ cl for normal hierarchy. Red - dot-dashed curves for THTO and blue - dotted curves for THWO}
 
\begin{minipage}{14pc}
\includegraphics[width=18pc]{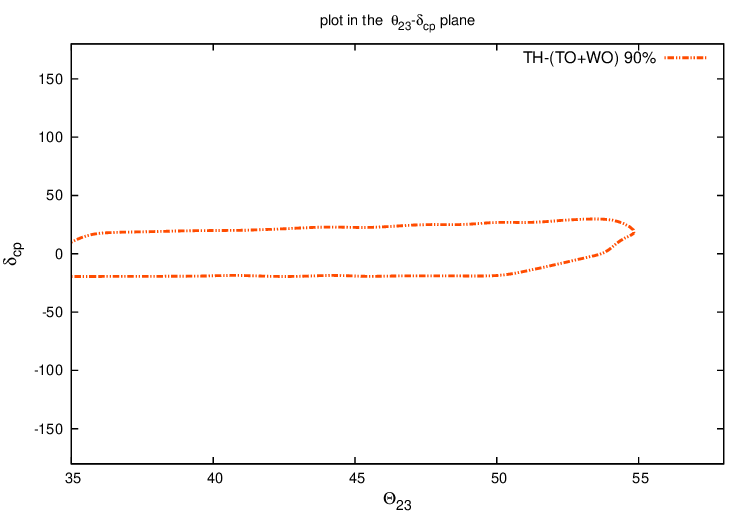}
\end{minipage}\hspace{2.5pc}%
\begin{minipage}{14pc}
\includegraphics[width=18pc]{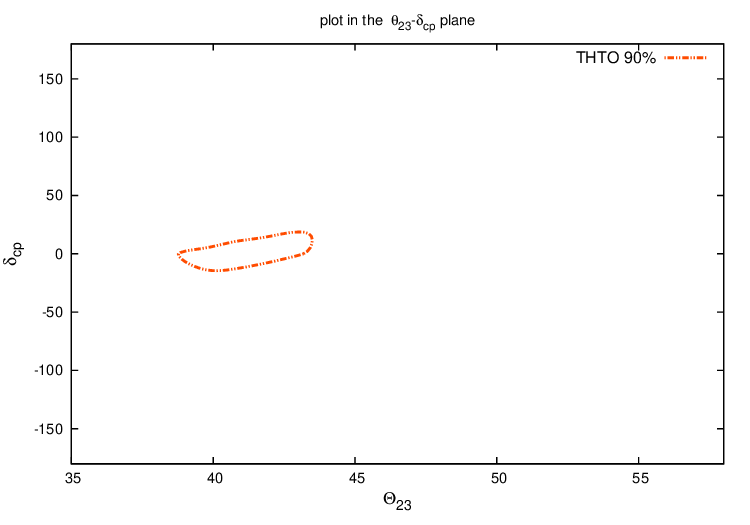}
\end{minipage} 
\caption{Contours in $\delta_{CP}-\theta_{23}$ plane for LBNE-like (1st fig.) and LBNE-like+Daya Bay setup (2nd fig.) at 90$\%$ cl at $\delta_{cp}=0$.}
 
\end{figure}
Using above details of LBNE experiment, we have produced exclusion
curves in the $\delta_{CP}-\theta_{13}$ plane. The best fit values
of oscillation parameters \cite{13} used
are: $\sin^{2}\theta_{12}=0.307$, $\sin^{2}\theta_{23}=0.427$,$\sin^{2}\theta_{13}=0.0241$
 and 
$\Delta{m}_{21}^{2}=7.54\times10^{-5}eV^{2}$ and 
$\Delta{m}_{31}^{2}=2.43\times10^{-3}eV^{2}$.\\
 Energy resolution used is 10$\%$ for the electrons,
and 5$\%$ for the muons. 1$\%$ systematic errors for the signal,
and 5$\%$ systematic error for the background has been used.
While searching for degenerate curves, we give the global best-fit values (true hierarchy, true octant) as true values. Then a chi-square analysis is done, to
search for true hierarchy, and wrong octant for the minima of $\chi^2$ function. We have used the $P_{\mu{e}}$ channel in our calculation and we have marginalized over $\delta_{CP}$, $sin^22\theta_{13}$, $\Delta{m}_{31}^{2}$ and $\theta_{23}$ in their $3\sigma$ ranges. We are only showing plots for normal hierarchy cases.
 In the plots we abbreviated
true hierarchy- true octant as THTO, true hierarchy- wrong octant as
THWO.\\
From these figures, following observations are in order :
\begin{enumerate}
\item Appearance of THTO and THWO (true hierarchy and wrong octant) contours in fig.1 (1st column), indicates the presence of octant degeneracy in the LBNE like experimental setup.
\item It may be noted that the regions inside
the contours are the allowed regions of the neutrino oscillation parameters
$\theta_{13}$ and $\delta_{CP}$.
\item Occurrence of THTO and THWO curves nearly at same $\delta_{cp}$ affects the measurement of $\theta_{13}$ but not of $\delta_{cp}$. It means that we can
pinpoint the value of $\delta_{cp}$ even in the presence of octant
degeneracy.

\item In figure 1 (2nd column), we see that THWO curves disappear, i.e., the Octant
degeneracy has been resolved completely, after combining LBNE with Daya
Bay experiment. Also we find that the size of the allowed regions
decrease as compared to figs 1 (1st column), i.e. the measurements of the
neutrino oscillation parameters $\theta_{13}$ and $\delta_{CP}$
become more deterministic. In our opinion, this resolution of Octant
degeneracy is because of the fact that, the difference in the values
of $\theta_{13}$, for the true and Octant degenerate curves,
is more than the sensitivity of the Daya Bay reactor experiment. And
therefore, the reactor data can pick up the true value of $\theta_{23}$.

\item In fig.2 (1st  fig.) the octant sensitivity is shown in $\delta_{CP}-\theta_{23}$ plane for LBNE like setup at $\delta_{CP}=0$ degree . It is seen that the contour is spanning over both the octant (true and wrong). Hence from this contour, it is not possible to figure it out which one is the correct octant. But in fig.2 (2nd fig.), which is for the LBNE-like+Daya Bay setup, ambiguity of octant is resolved. From fig.2 (2nd fig.), it is clear that lower octant ($\theta_{23}< 45^0$) is the true octant as indicted by the disappearance of allowed region in the wrong octant or higher octant of $\theta_{23}$. We have checked the results for different $\delta_{CP}$ and the results are consistent. Since, lower octant is also preferred by Global fit data, with the help of this proposal, we have been able to pinpoint the octant in LBNE which is the highlight of this work. 
\end{enumerate}

\section{Discussion and Conclusion}
To conclude, in this work, we have presented interesting results on resolution
of Octant degeneracy in LBNE-like setup.
The Octant degeneracy is resolved after combining LBNE-like with Daya
Bay reactor experiment. From Fig.2, we can conclude that, in presence Daya Bay reactor setup can pinpoint the true octant of LBNE like setup. Since Octant degeneracy
poses a major hurdle to precise measurement of neutrino oscillation
parameter $\theta_{23}$ at long baseline neutrino experiments, results presented
in this work are extremely important and relevant.

\section{Acknowledgement}
KB would like to thank Thomas Schwetz, Raj Gandhi and Pomita Ghoshal for fruitful discussions. She also would like
to thank UGC-SAP program, and HRI (Harish Chandra Research Institute),
Allahabad, for financial support, to visit HRI. DD would like to thank HRI, Allahbad
for financial support to visit the Institute.

\end{document}